\documentclass[sigconf]{acmart}

\AtBeginDocument{%
  }

\setcopyright{cc}
\setcctype{by-nc-nd}
\copyrightyear{2026}
\acmYear{2026}
\acmBooktitle{}
\acmDOI{}
\acmISBN{}

\acmConference[arXiv 2026]{arXiv}{May 2026}{}

\begin{document}
\title{SmellBench: Evaluating LLM Agents on Architectural Code Smell Repair}
\author{Ion George Dinu}
\email{dinu.george0019@gmail.com}
\affiliation{%
  \institution{University of Craiova}
  \city{Craiova}
  \country{Romania}}

\author{Marian Cristian Mih\u{a}escu}
\email{cristian.mihaescu@edu.ucv.ro}
\affiliation{%
  \institution{University of Craiova}
  \city{Craiova}
  \country{Romania}}

\author{Traian Rebedea}
\email{traian.rebedea@upb.ro}
\affiliation{%
  \institution{NVIDIA}
  \institution{University Politehnica of Bucharest}
  \city{Bucharest}
  \country{Romania}}

\begin{abstract}

Architectural code smells erode software maintainability and are costly to repair manually, yet unlike localized bugs, they require cross-module reasoning about design intent that challenges both developers and automated tools.
While large language model agents excel at bug fixing and code-level refactoring, their ability to repair architectural code smells remains unexplored.
We present the first empirical evaluation of LLM agents on architectural code smell repair. We contribute SmellBench, a task orchestration framework that incorporates smell-type-specific optimized prompts and supports iterative multi-step execution, together with a scoring methodology that separately evaluates repair effectiveness, false positive identification, and net codebase impact. We evaluate 11 agent configurations from four model families (GPT, Claude, Gemini, Mistral) on 65 hard-severity architectural smells detected by PyExamine in the Python project scikit-learn, validated against expert judgments.
Expert validation reveals that 63.1\% of detected smells are false positives, while the best agent achieves a 47.7\% resolution rate. Agents identify false positives with up to $\kappa = 0.94$ expert agreement, but repair aggressiveness and net codebase quality are inversely related: the most aggressive agent introduces 140 new smells.
These findings expose a gap between current LLM capabilities in localized code transformations and the architectural understanding needed for cross-module refactoring.
SmellBench provides reusable infrastructure for tracking progress on this underexplored dimension of automated software engineering. We release our code and data at \url{https://doi.org/10.5281/zenodo.19247588}.
 
\end{abstract}

\maketitle

\section{Introduction}
\label{sec:introduction}

As software systems grow in size and complexity, they accumulate structural degradations that increase the cost of maintenance, slow feature development, and raise the risk of introducing defects during changes. In practice, development teams address these issues through manual code reviews and periodic refactoring efforts, but these processes are time-consuming, require deep knowledge of the system's architecture, and are often deferred in favor of feature work, allowing the degradation to compound over time.
 
Code smells can be categorized by scope: implementation smells arise within methods, structural smells involve class-level design, and architectural smells manifest at the component level~\cite{suryanarayana2014refactoring}. An architectural smell is a structural decision at the package or module level that violates accepted design principles and negatively affects system-wide quality attributes such as maintainability, evolvability, and modularity~\cite{jolak2025maintainability}.
 
Static analysis tools can flag candidate smells~\cite{shivashankar2025pyexamine}, but they frequently report false positives, cases where the flagged pattern is actually an intentional design choice rather than a genuine defect~\cite{guo2023falsepositive}. Developers must therefore exercise architectural judgment to distinguish genuine issues from acceptable design patterns.
 
Large language model (LLM) agents have rapidly advanced automated software engineering. On localized bug repair, leading agents now resolve over 70\% of SWE-bench Verified\footnote{\url{https://www.swebench.com}} issues, and LLM-based approaches have been applied to code smell detection, method-level refactoring, and design issue localization (Section~\ref{sec:related_work}). However, these efforts operate at the method or class level and predominantly target Java codebases.
 
No prior work has evaluated LLM agents on the repair of architectural code smells. This gap exists for two reasons. First, no benchmark has been available for this task: existing benchmarks such as SWE-bench~\cite{jimenez2024swebench} target bug fixes where a failing test provides a clear oracle, but architectural smells have no pass/fail signal and require holistic assessment of design quality. Second, architectural smell repair demands capabilities that differ qualitatively from bug fixes. Where a bug fix targets a localized defect with a well-defined expected behavior, resolving an architectural smell may require coordinating changes across dozens of modules, reasoning about transitive dependency chains, and preserving design invariants that are nowhere explicitly documented.
 
We ask whether current LLM agents can effectively repair architectural code smells, and how their capabilities vary across repair effectiveness, false positive identification, handling of borderline cases where the detection is partially valid, and net codebase impact. We organize our methodology and experiments around four research questions:
 
\begin{description}
  \item[\textbf{RQ1}] How effective are LLM agents at repairing genuine architectural smells?
  \item[\textbf{RQ2}] How accurately do agents identify false positives?
  \item[\textbf{RQ3}] How do agents handle ambiguous, partially valid smells?
  \item[\textbf{RQ4}] What is the net impact of agent interventions on overall codebase quality?
\end{description}
 
To address these questions, we make three contributions. First, we present \textbf{SmellBench}, a task orchestration framework built on the Model Context Protocol~\cite{anthropic2024mcp} that pairs each architectural smell type with an optimized execution playbook and supports iterative multi-step repair through a structured task lifecycle. Second, we construct an \textbf{expert-validated benchmark} of 65 hard-severity architectural smells detected by PyExamine~\cite{shivashankar2025pyexamine} in the Python project scikit-learn~\cite{pedregosa2011scikit}. Third, we conduct the \textbf{first multi-agent empirical evaluation} on architectural smell repair, spanning 11 configurations from four model families using the leading coding agent CLIs available at the time of evaluation: OpenAI Codex, Claude Code, Gemini CLI, and Mistral Vibe.

\section{Related Work}
\label{sec:related_work}
 
\subsection{LLM-Based Program Repair}
 
Automated program repair (APR) has been transformed by large language models. Xia et al.~\cite{xia2023apr} established that pre-trained language models substantially outperform all existing APR techniques across five benchmark datasets spanning three programming languages. Subsequent work has explored multiple paradigms: RepairAgent~\cite{bouzenia2025repairagent} uses a finite state machine to constrain tool availability at each repair phase, SWE-agent~\cite{yang2024sweagent} combines LLMs with agent-computer interfaces for repository-level issue resolution, and Agentless~\cite{xia2025agentless} showed that a simple three-phase process of localization, repair, and patch validation can match complex agents, resolving 32.67\% of SWE-bench Lite~\cite{jimenez2024swebench} tasks.
 
All of these approaches rely on test-based oracles: a bug is fixed when a failing test passes. This binary signal also serves as a verifiable reward for reinforcement learning approaches that further improve agent performance. Architectural smells have no failing test, no binary pass/fail signal, and consequently no verifiable reward that can drive learned optimization. Furthermore, existing agents operate on localized code regions, whereas architectural smells span multiple modules and require reasoning about design intent, dependency structures, and framework conventions.
 
Beyond bug repair, LLM agents have also been applied to design quality tasks. Batole et al.~\cite{batole2025localizeagent} proposed LOCALIZEAGENT, a multi-agent framework that transforms program analysis outputs into LLM-friendly representations for design issue localization, achieving up to 206\% improvement over baseline prompting on tasks such as God Class and Cyclomatic Complexity detection. However, LOCALIZEAGENT focuses on localization (identifying which methods should be refactored) and does not generate repair code.

\subsection{Code Smell Detection and Refactoring with LLMs}
 
Recent work applies LLMs to code smell detection and refactoring, but exclusively at the class or method level and predominantly on Java codebases. Wu et al.~\cite{wu2024ismell} proposed iSMELL, combining LLMs with expert toolsets via a Mixture-of-Experts architecture for detecting God Class, Feature Envy, and Refused Bequest, achieving an average detection F1 of 75.17\%. Cordeiro et al.~\cite{cordeiro2024refactoring} confirmed that LLMs handle structurally simple refactorings well but struggle with complex, context-dependent design smells such as Broken Modularization. The only large-scale Python-focused work is SmellCC~\cite{xue2025smellcc}, which eliminated 91.6\% of over 200K smells in the CodeSearchNet-Python dataset; however, SmellCC targets method-level implementation smells (e.g., naming conventions, long parameter lists) rather than cross-module structural issues.
 
Multi-agent architectures have shown promise for refactoring. MANTRA~\cite{xu2025mantra} uses three agents with retrieval-augmented generation to achieve compilation and test pass rates comparable to human refactorings, and RefAgent~\cite{oueslati2025refagent} reduces code smells by a median of 52.5\% across eight Java projects using four specialized agents. Both operate at the class level within single files. At the multi-file level, RefactorBench~\cite{gautam2025refactorbench} provides 100 handcrafted tasks across 9 Python repositories, where the best agent solves only 22\% compared to 87\% for a human developer, confirming that multi-file reasoning remains a major bottleneck. While RefactorBench targets general code restructuring, no existing benchmark specifically addresses the repair of architectural smells.
 
A complementary direction examines smells introduced by LLMs. Velasco et al.~\cite{velasco2026causal} performed a causal analysis showing that prompt design and model architecture are the dominant factors influencing smell propensity during code generation, and that targeted mitigation strategies can reduce smell occurrence at inference time.

\subsection{Architectural Code Smell Repair}
 
Architectural smells have been extensively catalogued and shown to negatively impact software maintainability~\cite{jolak2025maintainability}, with systematic mappings of detection tools~\cite{mumtaz2021mapping} and recent Python-specific detectors like PyExamine~\cite{shivashankar2025pyexamine} now providing comprehensive coverage.
 
The closest work to ours is Tessa et al.~\cite{tessa2025ecsa}, who evaluated Gemini 1.5 Pro on detecting Hub-like Dependency smells across 39 Java projects (100 positive and 35 negative cases), achieving 100\% recall but variable precision (64--82\%) and only 49\% satisfactory explanations. This study covers a single smell type with a single model and does not attempt repair. ROSE~\cite{nursapa2025rose} uses fine-tuned CodeBERT and CodeT5 to classify which refactoring strategy applies to architectural smells, achieving up to 96.9\% accuracy, but classifies strategies rather than executing repairs. Both studies target Java exclusively, leaving Python architectural smells unstudied.
 
To the best of our knowledge, no prior work evaluates LLM agents on the repair of architectural smells, nor has any study quantified false positive rates for architectural smell detectors through expert validation, despite false positives exceeding 50\% in static analysis practice~\cite{guo2023falsepositive}. This gap persists because no benchmark existed for this task, and because the cross-module nature of these smells demands agent capabilities (multi-file reasoning, dependency-aware planning, and build validation) that only recently became feasible with autonomous coding agents~\cite{yang2024sweagent,bouzenia2025repairagent}.

\section{SmellBench - A Benchmark for Architectural Code Smells}
\label{sec:benchmark}
 
We target scikit-learn~\cite{pedregosa2011scikit} (v1.7.2\footnote{\url{https://github.com/scikit-learn/scikit-learn/tree/1.7.2}}), a widely-used Python machine learning library comprising approximately 200K lines of Python code (excluding tests) across 100+ modules. Its size, modular architecture, and active maintenance history make it a realistic subject for architectural smell analysis. We selected a Python codebase for two reasons. First, as established in Section~\ref{sec:related_work}, existing smell detection and refactoring research is almost exclusively Java-focused, leaving Python, the dominant language in the machine learning and data science ecosystem, entirely unstudied at the architectural level. Second, scikit-learn's extensive test suite and continuous integration infrastructure provide a natural validation oracle for assessing whether agent-produced refactorings preserve functional correctness.
 
\subsection{Smell Detection and Selection}
\label{sec:detection}
 
We run PyExamine~\cite{shivashankar2025pyexamine}, a comprehensive architectural smell detection tool for Python, on the scikit-learn codebase. PyExamine identifies smells through dependency analysis and software metrics, covering seven architectural smell types: Hub-like Dependency, Scattered Functionality, Cyclic Dependency, God Object, Unstable Dependency, Improper API Usage, and Redundant Abstractions.
 
PyExamine detects hundreds of smell instances across scikit-learn. To focus the evaluation on the most challenging cases, we apply a two-stage filtering process. First, we retain only the five smell types that have sufficient instances in scikit-learn for meaningful evaluation: Scattered Functionality, Potential Redundant Abstractions, Unstable Dependency, Potential Improper API Usage, and God Object. Hub-like Dependency and Cyclic Dependency are excluded due to insufficient instances in the target codebase.
 
Second, we classify each smell instance into three difficulty tiers (easy, medium, hard) using a deterministic, rule-based scheme. Each smell type maps to a single numeric metric extracted from the detection report that reflects the contextual complexity an agent must handle. Table~\ref{tab:thresholds} lists the metric and thresholds for each smell type.

\begin{table}[!htb]
\centering
\caption{Difficulty classification thresholds by smell type.}
\label{tab:thresholds}
\begin{tabular}{llccc}
\toprule
\textbf{Smell Type} & \textbf{Metric} & \textbf{Easy} & \textbf{Med.} & \textbf{Hard} \\
\midrule
Scattered Func.      & \# modules       & $\leq$4    & 5--7     & $\geq$8  \\
Redundant Abstr.     & similarity \%    & $<$70\%    & 70--89\% & $\geq$90\% \\
Unstable Dep.        & outgoing deps    & $\leq$8    & 9--19    & $\geq$20 \\
Improper API         & total repeats    & $\leq$40   & 41--129  & $\geq$130 \\
God Object           & public functions & $\leq$40   & 41--59   & $\geq$60 \\
\bottomrule
\end{tabular}
\end{table}

We select only the hard difficulty tier for our benchmark, as these instances affect core modules, cross multiple package boundaries, and demand the most sophisticated reasoning from agents. This yields 65 candidate smells: Scattered Functionality~(20), Potential Redundant Abstractions~(25), Unstable Dependency~(14), Potential Improper API Usage~(4), and God Object~(2).

 \begin{figure*}[t]
  \centering
  \includegraphics[width=0.75\textwidth]{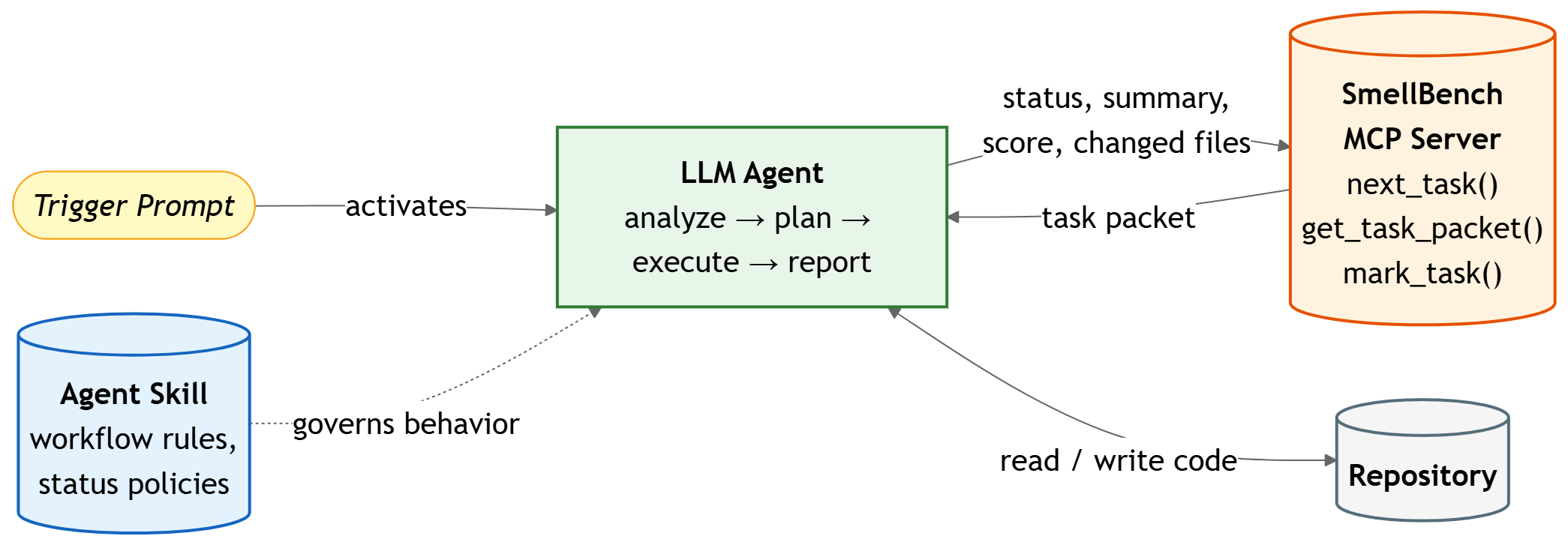}
  \caption{Overview of the SmellBench agent framework.}
  \label{fig:architecture}
\end{figure*}

\subsection{Expert Validation}
\label{sec:validation}
 
Static analysis tools are known to produce high false positive rates~\cite{guo2023falsepositive}, and architectural smell detectors are no exception. Expert validation proceeded in two phases.
 
In the first phase, one of the authors, a senior software engineer with architectural review experience, established ground truth by manually classifying each of the 65 detected smells. The expert reviewed the smell descriptions, the flagged source code, and scikit-learn's architectural patterns and design conventions, forming a preliminary verdict for each smell. The repair summaries from all 11 agents on all 65 tasks were then consulted to ensure no relevant code paths or design justifications were overlooked; no preliminary verdicts were changed as a result. Each smell was assigned one of three verdicts:
 
\emph{False Positive} (41 smells, 63.1\%): the detected pattern is an intentional or acceptable design choice. Common justifications include the Template Method pattern (where each subclass legitimately overrides a method), framework-mandated interface implementations, and deliberate facade modules that aggregate public APIs.
 
\emph{True Positive} (11 smells, 16.9\%): the smell is a genuine architectural issue that warrants repair. These include redundant implementations that could be consolidated and unnecessary coupling between modules.
 
\emph{Partially Valid} (13 smells, 20.0\%): the detection has merit, but the severity is overstated or the design trade-off is debatable. Partial improvement is possible and beneficial, but full elimination would conflict with established design constraints.
 
In the second phase, the expert reviewed the actual code changes produced by all 11 agents across all 65 tasks (715 agent-task outcomes) to verify that reported repairs correspond to genuine architectural restructuring rather than superficial transformations that merely satisfy the detection tool. Repair success itself is measured objectively by re-running PyExamine on each post-fix codebase (Section~\ref{sec:methodology}), not by agent self-reports.

\begin{table}[!htb]
\centering
\caption{SmellBench benchmark composition: expert verdicts by smell type.}
\label{tab:benchmark}
\begin{tabular}{lcccc}
\toprule
\textbf{Smell Type} & \textbf{Total} & \textbf{FP} & \textbf{TP} & \textbf{Partial} \\
\midrule
Scattered Functionality        & 20 & 13 &  2 &  5 \\
Redundant Abstractions       & 25 & 25 &  0 &  0 \\
Unstable Dependency          & 14 &  2 &  6 &  6 \\
Improper API Usage           &  4 &  0 &  2 &  2 \\
God Object             &  2 &  1 &  1 &  0 \\
\midrule
\textbf{Total}         & \textbf{65} & \textbf{41} & \textbf{11} & \textbf{13} \\
\bottomrule
\end{tabular}
\begin{flushleft}
\footnotesize
FP = False Positive (intentional design). TP = True Positive (genuine smell). Partial = partially valid detection.
\end{flushleft}
\end{table}

Table~\ref{tab:benchmark} summarizes the distribution of expert verdicts across smell types. The high overall false positive rate (63.1\%) is consistent with prior findings on static analysis~\cite{guo2023falsepositive} and varies substantially across smell types: Potential Redundant Abstractions are entirely false positives (25/25), while Unstable Dependencies are predominantly genuine issues (6/14 True Positive, 6/14 Partially Valid). Although the Redundant Abstractions instances contain no true positives, we retain them in the benchmark because agent refactorings applied to other tasks can incidentally resolve these smells through cascading side effects, making them useful for measuring unintended cross-task impact (Section~\ref{sec:rq1}). This variation makes the benchmark a demanding test of agent judgment: a successful agent must not only repair confirmed smells but also correctly identify false positives.

To assess the reliability of the expert verdicts, two additional senior developers independently classified all 65 smells using the same three-category scheme. Because the categories are ordinal (False Positive, Partially Valid, True Positive), we measure agreement using quadratic-weighted Cohen's $\kappa$, which penalizes adjacent disagreements lightly (0.25) and extreme disagreements fully (1.00). The mean pairwise $\kappa_w = 0.67$ (substantial agreement), with all three pairs above 0.62. Disagreements concentrate on the False Positive / Partially Valid boundary; no annotator pair disagrees across the full scale (False Positive vs.\ True Positive) on more than 5 of 65 cases. As a complementary check, we collapse the categories into a binary classification (genuine vs.\ not genuine, merging False Positive and Partially Valid): annotators agree on 81.5\% of cases (Fleiss' $\kappa = 0.45$, moderate, on a heavily imbalanced distribution). The primary annotator's verdicts are used throughout the evaluation.

\subsection{Benchmark Design Rationale}
\label{sec:rationale}
 
The three-category expert validation directly enables evaluation of three distinct agent capabilities: repair effectiveness on the 11 true positives, false positive identification on the 41 false positives, and partial smell handling on the 13 ambiguous cases. This tripartite structure reflects the practical reality that in any real codebase, most tool-flagged smells will be false positives, and the partially valid category tests whether agents can make targeted improvements without over-correcting.

\section{Proposed Approach}

Figure~\ref{fig:architecture} shows the overall architecture. A trigger prompt activates an LLM agent whose behavior is governed by an agent skill, a structured prompt encoding workflow rules and status policies. The agent interacts with the SmellBench MCP server to retrieve task specifications and report outcomes, while reading and writing code directly in the target repository. The following subsections describe each component.

\subsection{Task Orchestration via MCP}
\label{sec:mcp}
 
SmellBench orchestrates agent-task interactions through a Model Context Protocol~\cite{anthropic2024mcp} (MCP) server that exposes three tool endpoints implementing a pull-based task scheduling loop. \texttt{next\_task} claims the next available task by scanning the task repository and atomically leasing it to the requesting agent. \texttt{mark\_task} records the agent's outcome, validating the payload and preserving the full history across retries. \texttt{get\_task\_packet} assembles the task packet that the agent uses to analyze and execute the task.

The task packet returned by \texttt{get\_task\_packet} is the central interface between the benchmark and the agent, functioning as a retrieval-augmented context assembled from multiple sources for each task. Its base composition includes the smell type, smell name, the detection report generated by PyExamine (describing the smell instance, affected entities, and severity metrics), the list of affected source files (deduplicated and normalized to relative paths), and the modules involved in the smell.

This structured context is augmented with a smell-specific execution playbook and three few-shot demonstrations, both produced by the GEPA prompt optimization pipeline (\S\ref{sec:gepa}). The playbook provides step-by-step refactoring instructions tailored to the smell type, while the demonstrations illustrate expected agent behavior for each terminal status (Done, Accepted, Need More Work). When a task was previously marked as Need More Work, the packet is further augmented with the agent's prior status, summary, and list of changed files, enabling the agent to continue incrementally from its previous progress rather than restarting from scratch.

\subsection{Task Lifecycle}
\label{sec:lifecycle}

\begin{figure}[!htb]
  \centering
  \includegraphics[width=\columnwidth]{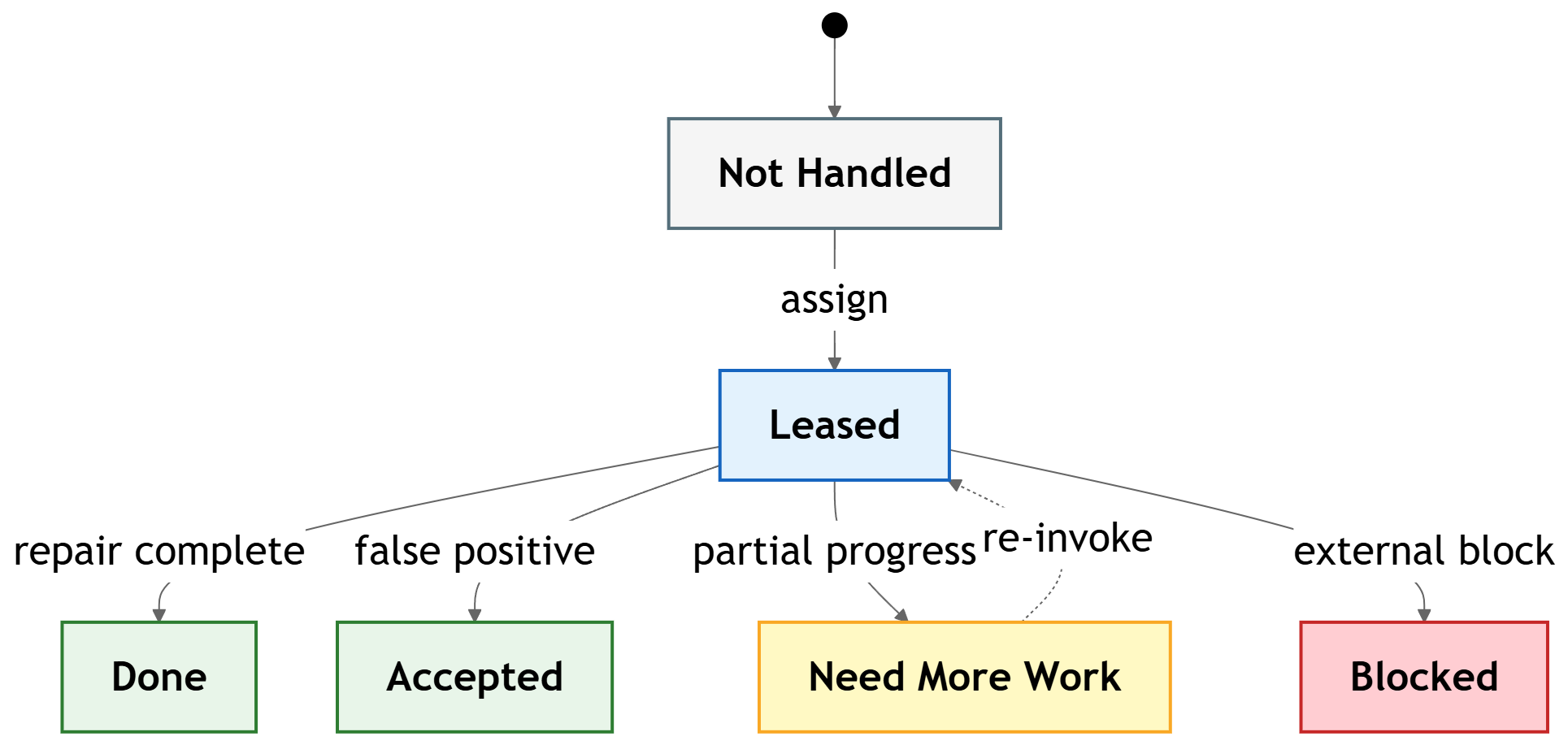}
  \caption{Task lifecycle state machine.}
  \label{fig:statemachine}
\end{figure}

Each task follows a structured lifecycle as shown in Figure~\ref{fig:statemachine}. Tasks begin in a $Not Handled$ state and transition to $Leased$ when an agent claims them via \texttt{next\_task}. The agent then follows a two-phase decision process.

In the first phase, the agent always attempts to solve the task. It analyzes the task packet, formulates a plan, and chooses one of two completion paths: single-iteration completion, where the plan is fully executable in one run, or multi-iteration completion, where the plan is too large and must be split into concrete mini-tasks, with at least one completed in the current run.

Before selecting a terminal status, the agent runs a validation gate on all modified files: a full compilation check and an import smoke test that verifies all affected modules load without errors. If either check fails, the agent must fix the issues before proceeding. This ensures that every terminal status represents a syntactically and structurally valid codebase state.

In the second phase, the agent selects a terminal status based on the outcome of its attempt:
 
\emph{Done}: the smell metric is reduced as far as possible, the validation gate passes, and no further concrete reduction path remains. The agent must report baseline and final metric values with a rationale for the stopping condition.
 
\emph{Accepted}: the agent attempted a repair and determined that the smell is a clear false positive, i.e., the detected pattern reflects intentional design and no reduction is possible. The agent must provide concrete evidence, such as a baseline measurement proving why the smell cannot be reduced. Any code changes made during the attempt are reverted, preserving the original codebase. If any plausible refactoring path remains, the agent must use Need More Work instead.
 
\emph{Need More Work}: the metric is not yet fully reduced and another iteration is needed after partial progress. The agent must have completed at least one concrete mini-task in the current run and must list the remaining mini-tasks as explicit next steps. On re-invocation, the task packet includes the previous summary and changed files, enabling incremental continuation. There is no hard cap on iterations; the agent self-terminates when it determines that no further concrete progress can be made, at which point it must select Done or Accepted based on the evidence gathered.
 
\emph{Blocked}: external constraints (missing files, tool failures) prevent progress. The agent must document the specific blocker and any attempted workaround.

\subsection{Prompt Optimization via GEPA}
\label{sec:gepa}

\begin{figure}[!htb]
  \centering
  \includegraphics[width=\columnwidth]{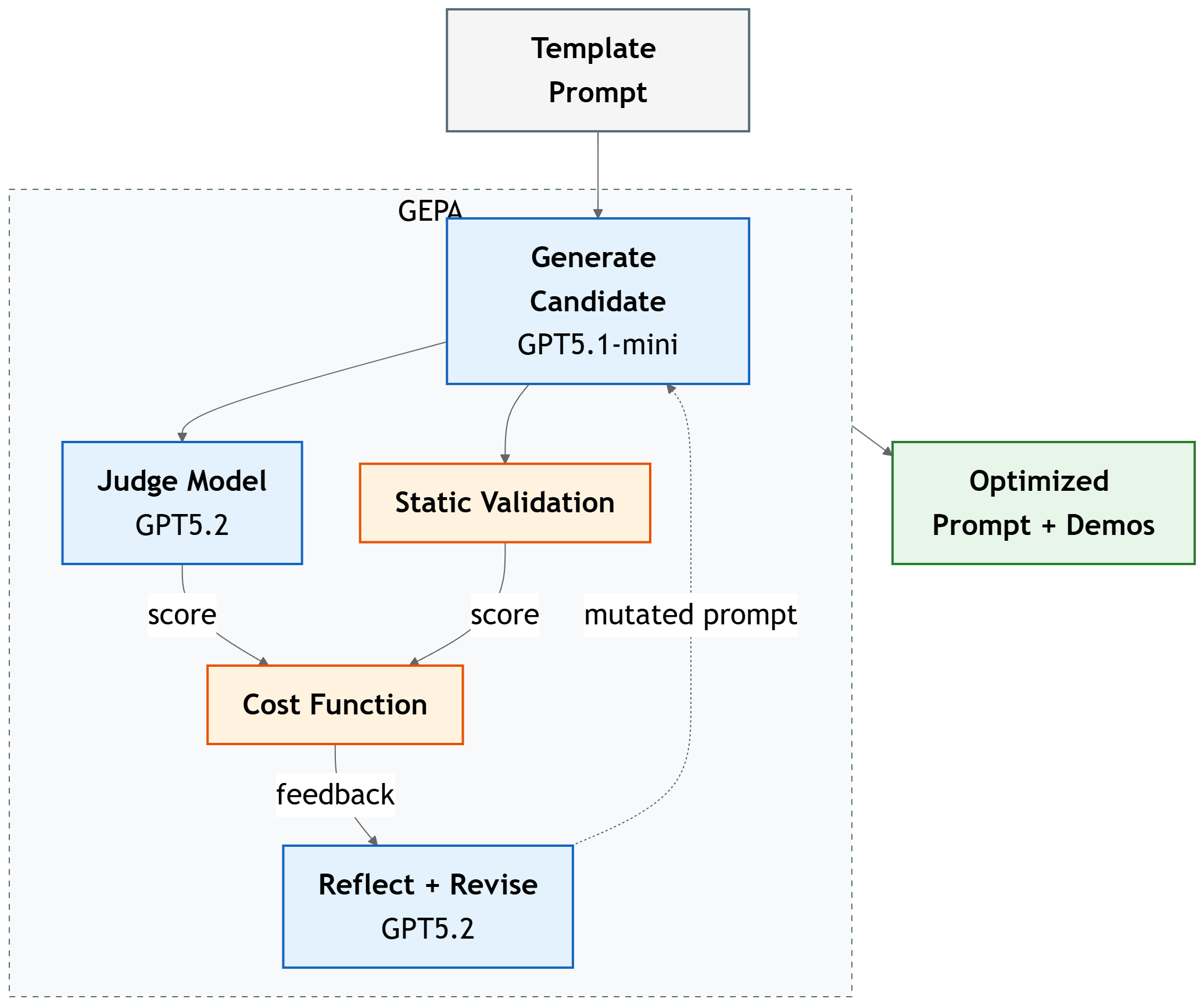}
  \caption{GEPA prompt optimization pipeline.}
  \label{fig:gepa}
\end{figure}

Rather than relying on a single hand-crafted prompt, each architectural smell type receives its own optimized execution playbook and few-shot demonstrations. We use GEPA~\cite{agrawal2025gepa}, a reflective prompt evolution optimizer, to automatically transform human-authored baseline templates into high-quality, smell-specific task packets (Figure~\ref{fig:gepa}).

\paragraph{Configuration.} For each smell type, we configure GEPA with four inputs: a human-authored baseline template as a warm start, a status policy encoding the decision rules from the task lifecycle (\S\ref{sec:lifecycle}), a strict output schema defining the required structure for the playbook and demonstrations, and training examples sampled from the easy and medium difficulty tiers (70/30 train-validation split). Although training used smells from the same codebase (scikit-learn), no hard-tier instances were included, ensuring the benchmark evaluation set remains unseen during optimization. The output is a finalized execution playbook and three few-shot demonstrations, one per terminal status, that are delivered to agents via the task packet (\S\ref{sec:mcp}).

\paragraph{Scoring function design.} GEPA requires a scoring function that evaluates candidate prompts at each iteration. The natural approach would be to deploy each candidate with an actual agent, measure task outcomes on the benchmark, and use the results as the score. However, a single agent run across all 65 tasks takes several hours, making end-to-end evaluation at each iteration computationally infeasible. We therefore designed a compound proxy metric combining two complementary signals.
 
A static validator (30\% weight) performs deterministic structural checks on the generated prompt and demonstrations. Through iterative development, we identified properties that consistently correlate with higher agent performance: explicit evidence requirements (baseline and final metric reporting), status labels with gating discipline, anti-loop rules for Need More Work, continuation logic for resumed tasks, and the absence of overly broad test commands or vague caution language. The validator also enforces demo quality: exactly three demonstrations, each with an explicit evidence chain. This component runs in milliseconds and filters structurally invalid candidates before the more expensive LLM evaluation.
 
A judge model (70\% weight) evaluates each candidate against a five-dimension rubric: goal alignment, execution quality, status discipline, demo quality, and output discipline. The judge returns both a normalized score and free-text feedback. The textual feedback is the key design choice: GEPA uses it to diagnose specific weaknesses and propose targeted improvements in subsequent iterations, enabling convergence with fewer candidates than scalar-only optimization.
 
\paragraph{Model selection.} We use a cost-efficiency-driven asymmetric architecture: a fast, smaller model (GPT-5.1-mini) for candidate generation, where throughput matters because many candidates are sampled per iteration, and a more capable model (GPT-5.2) for both judging and reflection, where reasoning quality matters because these roles must evaluate prompt quality against a multi-dimensional rubric and diagnose failures to propose concrete improvements. GEPA is configured in its light preset, which prioritizes quick convergence over exhaustive search.

\paragraph{Necessity of prompt optimization.} During development, we attempted to use hand-crafted prompts that described the refactoring goal without the structured playbook and few-shot demonstrations. These prompts consistently failed: agents entered infinite Need More Work loops, left tasks in unhandled states, or produced outputs that did not conform to the status protocol. GEPA-optimized prompts resolved these failure modes by encoding explicit anti-loop rules, status gating discipline, and structured evidence requirements. The baseline comparison (Section~\ref{sec:methodology}) further confirms that even a capable model (Claude~Haiku~4.5) achieves substantially lower performance without the full task protocol.
 
\subsection{Agent Architecture}
\label{sec:agent}

We evaluate off-the-shelf coding agent CLIs from four provider families: OpenAI Codex~\cite{openai2025codex}, Claude Code~\cite{anthropic2025claude}, Gemini CLI~\cite{google2025gemini}, and Mistral Vibe~\cite{mistral2025vibe}. All agents receive the same agent skill, a hand-written instruction set that encodes the SmellBench protocol: the MCP call sequence, the status taxonomy with its evidence requirements (\S\ref{sec:lifecycle}), and the completion strategy for multi-iteration tasks. The skill defines the workflow (claim task, execute, report), status selection rules with anti-loop safeguards, evidence requirements for each terminal status, and mandatory tool usage constraints. The full agent skill is included in our replication package. Each agent invocation is stateless: the agent carries no memory across runs, and all task persistence is handled by the MCP server, making the task packet the sole source of context for each execution. The agent operates autonomously, without human intervention.

\section{Experimental Results}
\label{sec:results}

\subsection{Evaluation Methodology}
\label{sec:methodology}

Using the four agent CLIs described in Section~\ref{sec:agent}, we deploy 10 model configurations across four provider families: GPT~5.4, GPT~5.3~Codex, GPT~5.3~Codex~Spark, GPT~5.1~Codex~Max (via OpenAI Codex); Claude~Opus~4.6, Claude~Sonnet~4.6, Claude~Haiku~4.5 (via Claude Code); Gemini~3.1~Pro~Preview, Gemini~3.1~Flash~Lite~Preview (via Gemini CLI); Devstral~v2 (via Mistral Vibe). All agents use default reasoning settings as provided by their respective CLIs, with no extended thinking or high-effort reasoning modes enabled. We also include a baseline: Claude~Haiku~4.5 run via Claude Code with only a minimal prompt listing the detected smell and affected files, without agent skill, status lifecycle, or GEPA-optimized task packet.

To verify functional correctness, we run scikit-learn's full test suite on each post-fix codebase. No test deviations from the original codebase were observed for any agent configuration, confirming that all refactorings preserve functional correctness.
 
Each agent is evaluated on all 65 hard-severity smells from the benchmark (Section~\ref{sec:benchmark}). Unlike bug repair, where a failing test provides a binary oracle, architectural smell repair lacks a clear pass/fail signal: agents either attempt a fix (with outcomes ranging from full repair to worsening) or flag the smell as a false positive. To capture this spectrum, we define a per-task scoring function $s \in [-1, 1]$ based on the agent's action and the objective outcome (Table~\ref{tab:scoring}).

\begin{table}[!htb]
\centering
\caption{Per-task scoring function $s \in [-1, 1]$.}
\label{tab:scoring}
\begin{tabular}{llc}
\toprule
\textbf{Agent Action} & \textbf{Outcome} & $\boldsymbol{s}$ \\
\midrule
Attempted fix & Fully repaired    & $1.0$ \\
Attempted fix & Partially improved & $(0, 1)$\textsuperscript{$\dagger$} \\
Attempted fix & Unchanged          & $0.0$ \\
Attempted fix & Worsened           & $[-1, 0)$\textsuperscript{$\dagger$} \\
\midrule
Flagged FP    & Expert: FP               & $1.0$ \\
Flagged FP    & Expert: Partially Valid  & $0.0$ \\
Flagged FP    & Expert: TP               & $-1.0$ \\
\bottomrule
\end{tabular}
\begin{flushleft}
\footnotesize
\textsuperscript{$\dagger$}Proportional to severity change: $s = -\Delta_{\text{sev}} / \text{sev}_{\text{old}}$, capped at $[-1, 1]$. FP = False Positive. TP = True Positive.
\end{flushleft}
\end{table}

Per-task scores are averaged over each expert category as described in Section~\ref{sec:rationale} to obtain three component metrics: True Positive Component (average $s$ score on $n{=}11$ genuine smells), False Positive Component ($n{=}41$ false positives), and Partial Component ($n{=}13$ ambiguous cases). These are combined into a single Weighted Effectiveness score that can be used to easily compare different agents and LLMs on the SmellBench benchmark:
 
\begin{equation}
\label{eq:weighted}
E = 0.60 \times \text{TP} + 0.20 \times \text{FP} + 0.20 \times \text{Partial}
\end{equation}
 
True Positive receives 60\% weight as the primary and most demanding capability; false positive identification and partial handling each receive 20\% as equally important judgment tasks.

\subsection{Overall Results}
\label{sec:overall}

Table~\ref{tab:leaderboard} ranks all 11 agents by weighted effectiveness ($E$, Equation~\ref{eq:weighted}) and its three component scores. Three performance tiers emerge. The top tier (ranks 1--4: GPT~5.3~Codex~Spark, GPT~5.3~Codex, Gemini~3.1~Pro, GPT~5.4) achieves $E$ between 0.428 and 0.478, driven by strong True Positive Components ($\geq 0.357$). The middle tier (ranks 5--8: the three Claude agents and GPT~5.1~Codex~Max) clusters between $E = 0.347$ and 0.379, with higher False Positive Components but lower True Positive scores. The bottom tier includes the baseline ($E = 0.152$), Devstral~v2 ($E = 0.067$), and Gemini~3.1~Flash~Lite ($E = -0.177$); the latter two receive negative True Positive Components because they incorrectly flag genuine smells as false positives.

Among the top eight agents, the highest-ranked on each component differs: GPT~5.3~Codex~Spark leads on $E$ and Partial, GPT~5.3 Codex on True Positive, and Gemini~3.1~Pro on False Positive. No single agent dominates all dimensions, indicating that repair aggressiveness and false positive judgment represent competing strengths.

Bootstrap confidence intervals (10{,}000 resamples) confirm that tier-level differentiation is robust: the top-tier composite CIs do not overlap with the baseline (e.g., GPT~5.3~Codex~Spark $E$: $[0.24, 0.70]$ vs.\ baseline: $[0.07, 0.27]$). Within tiers, however, the True Positive and Partial Components carry substantial uncertainty due to small sample sizes ($n{=}11$ and $n{=}13$), with CI widths exceeding 0.50 and all top-eight True Positive CIs overlapping heavily. The False Positive Component is more precisely estimated ($n{=}41$; CI widths 0.12--0.24). Component-level rankings within a tier should therefore be interpreted with caution.

\begin{table}[t]
\centering
\caption{Overall agent ranking by weighted effectiveness.}
\label{tab:leaderboard}
\small
\begin{tabular}{rlcccc}
\toprule
 & \textbf{Agent} & $\boldsymbol{E}$ & \textbf{TP} & \textbf{FP} & \textbf{Partial} \\
 &                &                  & \footnotesize{($n{=}11$)} & \footnotesize{($n{=}41$)} & \footnotesize{($n{=}13$)} \\
\midrule
1 & GPT 5.3 Codex Spark        & \textbf{0.478} & 0.383          & 0.891          & \textbf{0.353} \\
2 & GPT 5.3 Codex        & 0.444          & \textbf{0.402} & 0.865          & 0.147          \\
3 & Gemini 3.1 Pro       & 0.435          & 0.389          & \textbf{0.952} & 0.056          \\
4 & GPT 5.4               & 0.428          & 0.357          & 0.909          & 0.158          \\
5 & Claude Haiku 4.5      & 0.379          & 0.300          & 0.951          & 0.042          \\
6 & Claude Opus 4.6       & 0.370          & 0.285          & 0.905          & 0.090          \\
7 & Claude Sonnet 4.6     & 0.354          & 0.278          & 0.881          & 0.056          \\
8 & GPT 5.1 Codex Max           & 0.347          & 0.304          & 0.786          & 0.037          \\
9  & Claude Haiku 4.5$^{b}$      & 0.152          & 0.099          & 0.449          & 0.013          \\
10 & Devstral v2          & 0.067          & $-$0.176       & 0.855          & 0.007          \\
11 & Gemini 3.1 Flash Lite    & $-$0.177       & $-$0.627       & 1.000          & 0.000          \\
\bottomrule
\end{tabular}
\begin{flushleft}
\footnotesize
TP = True Positive Component. FP = False Positive Component. $^{b}$Baseline: bare Claude Haiku 4.5 without agent skill or task protocol (Section~\ref{sec:methodology}).
\end{flushleft}
\end{table}

\subsection{RQ1: Repair Effectiveness}
\label{sec:rq1}

Table~\ref{tab:smell_type} breaks down resolution counts by smell type. Four of the ten non-baseline agents (all three Claude models and GPT 5.1 Codex Max) converge at exactly 18 resolved smells (27.7\%), suggesting a common floor for straightforward fixes. Differentiation occurs beyond this floor: GPT~5.3~Codex~Spark resolves 31 (47.7\%), while GPT~5.4, GPT~5.3 Codex, and Gemini~3.1~Pro resolve 21, 20, and 19 respectively. Not all resolutions result from the agent's work on the corresponding task: some smells disappear as cascading side effects of refactorings applied to other tasks. Section~\ref{sec:rq4} quantifies this distinction.

Partial severity reductions are common even when full resolution is not achieved. The partial improvement rate (PIR), defined as the percentage of attempted-but-unresolved smells where the severity metric decreased, ranges from 60.0\% (Claude~Haiku~4.5) to 82.4\% (GPT~5.3~Codex~Spark) among the top eight, versus only 21.3\% for the baseline despite the same total resolution rate, confirming that the agent skill substantially improves repair quality beyond what a bare prompt achieves.
 
Cross-referencing with expert verdicts (Table~\ref{tab:benchmark}) provides further context. For Scattered Functionality, 13 of 20 instances are false positives, leaving 7 genuinely fixable cases of which agents resolve 1 to 4 (14--57\%). All 25 Redundant Abstraction instances are false positives, yet agents still resolve 15 to 21 of them, because refactorings applied to other tasks can incidentally reduce the functional similarity scores that triggered these detections. Improper API Usage contains no false positives (2 genuine, 2 partially valid), making 0/4 resolution a genuine capability gap: these tasks require understanding API contracts and replacing repetitive call patterns with thin wrappers, a form of cross-function reasoning that no agent could perform. Unstable Dependency (12/14 genuine or partially valid) provides the sharpest differentiation: GPT~5.3~Codex~Spark resolves 7 through dependency restructuring, while all Claude agents and GPT~5.1~Codex~Max resolve none.

\begin{table}[!htb]
\centering
\caption{Smells resolved by type for all agents.}
\label{tab:smell_type}
\small
\begin{tabular}{lccccccc}
\toprule
\textbf{Agent} & \textbf{SF} & \textbf{RA} & \textbf{GO} & \textbf{IA} & \textbf{UD} & \textbf{Rate} & \textbf{PIR} \\
& \footnotesize{(20)} & \footnotesize{(25)} & \footnotesize{(2)} & \footnotesize{(4)} & \footnotesize{(14)} & \footnotesize{(\%)} & \footnotesize{(\%)} \\
\midrule
GPT 5.3 Codex Spark      & 3          & \textbf{21} & 0          & 0 & \textbf{7} & \textbf{47.7} & \textbf{82.4} \\
GPT 5.3 Codex      & 3          & 15          & \textbf{1} & 0 & 1          & 30.8 & 72.7 \\
Gemini 3.1 Pro     & 2          & 15          & \textbf{1} & 0 & 1          & 29.2 & 76.5 \\
GPT 5.4            & \textbf{4} & 15          & \textbf{1} & 0 & 1          & 32.3 & 69.6 \\
Claude Haiku 4.5   & 2          & 15          & \textbf{1} & 0 & 0          & 27.7 & 60.0 \\
Claude Opus 4.6    & 2          & 15          & \textbf{1} & 0 & 0          & 27.7 & 72.0 \\
Claude Sonnet 4.6  & 2          & 15          & \textbf{1} & 0 & 0          & 27.7 & 73.1 \\
GPT 5.1 Codex Max        & 2          & 15          & \textbf{1} & 0 & 0          & 27.7 & 81.5 \\
Claude Haiku 4.5$^{b}$    & 2          & 16          & 0          & 0 & 0          & 27.7 & 21.3 \\
Devstral v2        & 1          & 10          & 0          & 0 & 0          & 16.9 & 33.3 \\
Gem.\ 3.1 Flash Lite & 1        & 5           & 0          & 0 & 0          & 9.2  & 66.7  \\
\bottomrule
\end{tabular}
\begin{flushleft}
\footnotesize
SF = Scattered Functionality, RA = Redundant Abstractions, GO = God Object, IA = Improper API Usage, UD = Unstable Dependency. Rate = percentage of all 65 smells resolved. PIR = partial improvement rate on attempted-but-unresolved smells (percentage where severity decreased). $^{b}$Baseline.
\end{flushleft}
\end{table}
 
\textbf{Finding 1:} The best agent resolves 47.7\% of smells (31/65), while four agents share a 27.7\% floor (18/65). A substantial portion of resolutions are cascading side effects rather than direct fixes (Section~\ref{sec:rq4}). Improper API Usage resists all agents (0/4), and Unstable Dependency is the sharpest differentiator (0--7 resolved). The agent skill improves partial repair quality substantially (60--82\% PIR vs.\ 21\% baseline).

\subsection{RQ2: False Positive Identification}
\label{sec:rq2}
 
We evaluate false positive identification using Cohen's Kappa ($\kappa$) on the 52 unambiguous tasks (41 False Positive $+$ 11 True Positive), excluding the 13 partially valid cases where the correct classification is ambiguous. Precision measures how many of the agent's false positive claims are confirmed by the expert (partially valid claims count as 0.5), recall measures how many of the 41 confirmed false positives the agent flagged, and F1 is their harmonic mean. Table~\ref{tab:fp} presents the results.

\begin{table}[!htb]
\centering
\caption{False positive identification performance.}
\label{tab:fp}
\begin{tabular}{lccccc}
\toprule
\textbf{Agent} & \textbf{Flagged} & \textbf{Prec.} & \textbf{Rec.} & \textbf{F1} & $\boldsymbol{\kappa}$ \\
\midrule
Gemini 3.1 Pro     & 44 & 0.95          & \textbf{0.97} & \textbf{0.96} & \textbf{0.94} \\
Claude Haiku 4.5   & 47 & 0.92          & \textbf{0.97} & 0.95          & \textbf{0.94} \\
Claude Opus 4.6    & 37 & \textbf{1.00} & 0.90          & 0.95          & 0.80 \\
Claude Sonnet 4.6  & 36 & \textbf{1.00} & 0.88          & 0.94          & 0.75 \\
GPT 5.3 Codex      & 38 & 0.96          & 0.85          & 0.90          & 0.71 \\
GPT 5.4            & 36 & 0.97          & 0.83          & 0.90          & 0.67 \\
GPT 5.1 Codex Max        & 35 & 0.97          & 0.80          & 0.88          & 0.64 \\
GPT 5.3 Codex Spark      & 38 & 0.92          & 0.80          & 0.86          & 0.58 \\
Devstral v2        & 42 & 0.87          & 0.80          & 0.84          & 0.52 \\
Gemini 3.1 Flash Lite  & 61 & 0.77          & \textbf{1.00} & 0.87          & 0.37 \\
Claude Haiku 4.5$^{b}$ & 0 & ---       & 0.00          & ---           & 0.00 \\
\bottomrule
\end{tabular}
\begin{flushleft}
\footnotesize
Flagged = total tasks flagged as false positive (out of 65). Sorted by $\kappa$. $^{b}$Baseline.
\end{flushleft}
\end{table}
 
Gemini~3.1~Pro and Claude~Haiku~4.5 achieve the strongest expert agreement ($\kappa = 0.94$). Claude~Opus~4.6 and Claude~Sonnet~4.6 achieve perfect precision (1.00) at the cost of lower recall. Seven agents produce zero incorrect false positive claims (flagging a true positive as a false positive); this is not a by-product of conservative flagging, as GPT~5.3~Codex and GPT~5.4 flag 38 and 36 tasks, respectively while maintaining this property. Gemini~3.1~Flash~Lite flags 61 of 65 tasks (93.8\%) as false positives ($\kappa = 0.37$), incorrectly dismissing 8 genuine smells and 12 partially valid ones. The baseline, with no false positive capability, achieves $\kappa = 0.00$.

Examining how agents arrive at their false positive decisions reveals three distinct patterns across the 409 total claims. The majority (371, 90.7\%) are pure flags where the agent concludes false positive without attempting any code changes. A smaller set (30, 7.3\%) are partial-fix flags where the agent makes code changes before ultimately concluding that the smell reflects intentional design. The remaining 8 (2.0\%) are investigated flags where the agent iterates through Need More Work before concluding false positive. The dominance of pure flags indicates that most agents make their false positive judgments during initial analysis rather than through trial-and-error repair attempts.
 
\textbf{Finding 2:} False positive identification ranges from $\kappa = 0.00$ (baseline) to $\kappa = 0.94$ (Gemini~3.1~Pro, Claude~Haiku~4.5), with seven agents producing zero incorrect claims. The baseline ($\kappa = 0.00$) confirms that the agent protocol is necessary, while Flash~Lite's near-universal flagging ($\kappa = 0.37$) shows the protocol alone is insufficient without adequate model capability.
 
\subsection{RQ3: Handling Ambiguous Smells}
\label{sec:rq3}
 
The Partial Component proves the most discriminating dimension, with scores ranging from 0.353 (GPT~5.3~Codex~Spark) to 0.000 (Gemini~3.1~Flash~Lite). Agent strategies diverge sharply: Claude~Opus~4.6 and Claude~Sonnet~4.6 attempt all 13 fixes, while Gemini~3.1~Flash Lite flags 12 of 13 as false positives and achieves no improvement on the remaining one. GPT~5.3~Codex~Spark fully resolves 3 partially valid smells and GPT~5.4 resolves 1. Most agents achieve only marginal improvements, with severity reductions scoring between 0.0 and 0.1.

The wide score range reflects a fundamental strategy difference: agents that attempt repair on ambiguous cases can score positively through partial improvements, while agents that flag them as false positives score zero (neither rewarded nor penalized). The top-scoring agents on this component combine willingness to attempt with sufficient capability to reduce the smell metric, a combination that only GPT~5.3~Codex~Spark consistently achieves.
 
\textbf{Finding 3:} Ambiguous smell handling is the most discriminating dimension (score range 0.000--0.353). Most agents default to conservative strategies that yield marginal improvement; only GPT~5.3~Codex~Spark fully resolves 3 of 13 partially valid smells.

\subsection{RQ4: Net Codebase Impact}
\label{sec:rq4}
 
An agent that resolves targeted smells but creates new ones elsewhere may leave the codebase worse off. To quantify this, we re-run PyExamine on each post-fix codebase and compute the net smell delta: the number of smells that disappeared minus the number of new smells detected. Table~\ref{tab:net_impact} presents the results.
 
GPT~5.3~Codex~Spark resolves the most smells (31) but introduces 140 new ones (net $-$109), driven by module consolidations that create new dependency violations. Gemini~3.1~Pro and Claude~Opus~4.6 introduce only 3 each while resolving 19 and 18 (net $+$16 and $+$15). The baseline resolves 18 smells with only 4 introduced (net $+$14), but at the cost of no false positive identification.
 
Claude agents produce zero negative per-task scores across all tasks. Devstral~v2 and Gemini~3.1~Flash~Lite each worsen 8 tasks.
 
\begin{table}[!htb]
\centering
\caption{Net codebase impact. Sorted by net delta.}
\label{tab:net_impact}
\begin{tabular}{lcccc}
\toprule
\textbf{Agent} & \textbf{Resolved} & \textbf{Introd.} & \textbf{Net $\Delta$} & \textbf{Neg.} \\
\midrule
Gemini 3.1 Pro     & 19 & \textbf{3}   & \textbf{+16}  & 0 \\
Claude Opus 4.6    & 18 & \textbf{3}   & +15           & 0 \\
Claude Haiku 4.5$^{b}$  & 18 & 4            & +14           & 0 \\
Gemini 3.1 Flash Lite & 6 & 0            & +6            & 8 \\
Devstral v2        & 11 & 6            & +5            & 8 \\
Claude Haiku 4.5   & 18 & 16           & +2            & 0 \\
Claude Sonnet 4.6  & 18 & 26           & $-$8          & 0 \\
GPT 5.3 Codex      & 20 & 38           & $-$18         & 2 \\
GPT 5.4            & 21 & 40           & $-$19         & 0 \\
GPT 5.1 Codex Max        & 18 & 42           & $-$24         & 0 \\
GPT 5.3 Codex Spark      & \textbf{31} & 140 & $-$109        & 2 \\
\bottomrule
\end{tabular}
\begin{flushleft}
\footnotesize
Resolved = direct + cascading repairs (smells that disappeared from the post-fix report). Introd.\ = new smells detected after agent modifications. Net $\Delta$ = Resolved $-$ Introd. Neg.\ = tasks with negative per-task scores. $^{b}$Baseline.
\end{flushleft}
\end{table}
 
\textbf{Finding 4:} Repair aggressiveness and net codebase quality are inversely related. The best net deltas ($+$15, $+$16) are achieved by conservative agents that introduce $\leq$3 new smells; the most aggressive agent introduces 140.

\subsection{Ranking Robustness}
\label{sec:robustness}

To verify that our findings are not artifacts of the 60/20/20 weighting, we compute rankings under seven configurations: the primary weights, equal weights ($\frac{1}{3}$ each), repair-dominant (70/15/15), repair-only (100/0/0), false-positive-excluded (50/0/50), no-partial (50/50/0), and a simple unweighted average across all 65 tasks.
 
Kendall's $W = 0.939$ ($\chi^2 = 65.71$, $df = 10$, $p < 0.001$) indicates strong concordance. The top four agents (GPT~5.3~Codex~Spark, GPT~5.3~Codex, Gemini~3.1~Pro, GPT~5.4) consistently occupy ranks 1--4, the middle tier (Claude~Haiku, Opus, Sonnet, GPT~5.1 Codex Max) ranks 5--8, and the bottom three occupy ranks 9--11, all with only minor permutations across configurations.
 
To test whether the ranking differences are statistically meaningful, we conduct pairwise Wilcoxon signed-rank tests with Holm correction on per-task scores. Among the top eight agents, no pair differs significantly ($p > 0.05$ for all pairs), and effect sizes are negligible (Cliff's $|\delta| < 0.13$~\cite{arcuri2011practical}). In contrast, all top-eight agents differ significantly from the baseline with medium effect sizes ($p \leq 0.001$, $|\delta| \approx 0.41$--$0.47$). Devstral~v2 and Gemini~3.1~Flash~Lite do not differ significantly from the baseline ($p = 0.23$ and $p = 0.82$, respectively).
 
\textbf{Finding 5:} The three performance tiers (top eight, baseline, bottom two) are robust across seven weighting configurations ($W = 0.939$, $p < 0.001$). Within the top tier, no pairwise difference is statistically significant; practical differentiation arises from net codebase impact (RQ4) and false positive precision (RQ2).

\section{Discussion}
\label{sec:discussion}
 
\paragraph{The architectural reasoning gap.} The best agent's 47.7\% resolution rate contrasts with the 70\%+ rates on SWE-bench Verified~\cite{jimenez2024swebench}. While direct comparison is imprecise, as SWE-bench uses test-based pass/fail on localized bugs whereas SmellBench uses tool-measured severity reduction on cross-module smells, the smell-type breakdown (RQ1) reveals that the hardest cases, Improper API Usage and Unstable Dependency, require understanding API contracts and transitive coupling, confirming that architectural smell repair is a qualitatively different task from localized bug fixing. Given the documented impact of architectural smells on software maintainability~\cite{jolak2025maintainability}, closing this gap is a significant practical priority.
 
\paragraph{Representative failure patterns.} Examining agent trajectories on the two hardest smell types reveals distinct failure modes. For Improper API Usage (0/4 resolved), agents consistently recognize the repetitive call patterns but fail at the planning stage: they propose changes to individual call sites in isolation rather than designing a centralized abstraction, resulting in partial modifications that do not reduce the metric. For Unstable Dependency (0--7 resolved), agents that succeed restructure imports by moving shared utilities into stable base modules. Agents that fail typically identify the correct dependency chain but stop at analysis without executing the file moves and import rewrites needed to complete the refactoring, defaulting to a Need More Work status that never converges. These patterns suggest that the bottleneck is not architectural understanding but the ability to execute coordinated multi-file transformations that span module boundaries.
 
\paragraph{Cascading repair mechanisms.} A substantial portion of resolutions are cascading side effects: refactorings applied to one task incidentally resolve smells detected elsewhere, predominantly Redundant Abstractions. These cascading repairs occur when an agent consolidates duplicated logic in a shared module, which reduces functional similarity scores for other module pairs that referenced the same code. While cascading repairs represent genuine improvements to the codebase, they are not the result of targeted agent reasoning about the resolved smell, and their count is sensitive to task execution order.
 
\paragraph{False positive identification as a separable skill.} The contrast between strong false positive agreement ($\kappa \geq 0.58$ for all top-eight agents) and low True Positive Components ($\leq 0.402$) suggests these are separable capabilities. False positive identification involves pattern recognition, matching detected smells against known design patterns, which is well-suited to LLMs trained on large code corpora. Repair requires generative planning across module boundaries, with combinatorially larger search spaces. The 63.1\% false positive rate in our benchmark is consistent with the broader static analysis literature~\cite{guo2023falsepositive}, though the hard-tier selection may amplify this rate: harder instances affect core modules where intentional design patterns (e.g., Template Method, Facade Aggregation) are more prevalent.
 
\paragraph{Implications for deployment.} The inverse relationship between repair aggressiveness and net codebase quality (RQ4) echoes the well-documented patch overfitting phenomenon in automated program repair, where aggressive tools produce fixes that satisfy the oracle but degrade overall quality. Our findings extend this observation to architectural smell repair and LLM agents, and imply distinct deployment profiles: conservative agents suit autonomous CI pipelines, while aggressive agents suit human-supervised refactoring, where introduced smells can be reviewed. Future work could explore hybrid strategies that combine conservative false positive identification with aggressive but scoped repair. The specific model rankings reported here will inevitably shift as providers release new versions; the lasting contribution is the SmellBench infrastructure and evaluation methodology, which can be re-run on future agents to track progress.
 
\paragraph{Inter-agent agreement and calibration.} Fleiss' $\kappa$ across the 10 non-baseline agents is 0.531 (moderate agreement), with pairwise agreement clustering by model family (Claude: $\kappa = 0.58$--$0.78$; GPT: $\kappa = 0.49$--$0.78$) but also showing strong cross-family convergence (e.g., Gemini~3.1~Pro and Claude~Opus~4.6: $\kappa = 0.77$). Self-assessment calibration varies widely: Claude~Haiku~4.5 exhibits the highest correlation between self-assessed and actual scores ($r = 0.68$), while Gemini~3.1~Pro assigns identical confidence regardless of outcome ($r = 0.00$). Well-calibrated agents could flag uncertain repairs for human review, but most current models lack this metacognitive capability.

\section{Threats to Validity}
\label{sec:threats}
 
\paragraph{External validity.} Our evaluation targets a single Python project (scikit-learn) and one detection tool (PyExamine), so results may not generalize to other languages, frameworks, or detectors. The five evaluated smell types do not cover all possible architectural smells. We mitigate this by selecting the hard difficulty tier and evaluating across 11 diverse agent configurations from four model families. Extending the benchmark to additional projects is a priority for future work. Additionally, scikit-learn is one of the most popular Python repositories on GitHub and is included in several LLM training datasets, raising the possibility that agent performance benefits from memorization of scikit-learn's codebase rather than generalizable architectural reasoning. We note, however, that the 63.1\% false positive rate and the agents' inability to resolve Improper API Usage (0/4) suggest that memorization alone does not explain the observed results.
 
\paragraph{Internal validity.} Expert verdicts were assigned by a single primary annotator (one of the authors), which could introduce subjective bias. Multi-annotator validation with two additional senior developers yields substantial agreement ($\kappa_w = 0.67$, Section~\ref{sec:validation}), with disagreements concentrated on the False Positive / Partially Valid boundary rather than extreme misclassifications. LLM non-determinism may affect reproducibility; we did not run multiple trials per agent due to computational cost, though agent behavior patterns and tier-level rankings were consistent across all runs. Agent-produced refactorings could introduce functional regressions; we verify correctness by running scikit-learn's full test suite on each post-fix codebase (Section~\ref{sec:methodology}), with no deviations observed. Each agent is evaluated through its native CLI (OpenAI Codex, Claude Code, Gemini CLI, Mistral Vibe), which means observed differences reflect the combined effect of model capability and CLI scaffolding. We deliberately evaluate off-the-shelf practitioner tools as they ship, since this reflects real-world deployment conditions, but results should not be interpreted as isolated model comparisons.
 
\paragraph{Construct validity.} Our scoring function assigns equal weight to all smells within each category, regardless of individual severity. The True Positive ($n{=}11$) and Partial ($n{=}13$) Components have small sample sizes, so component-level rankings within a tier should be interpreted with caution (bootstrap confidence intervals are reported in Section~\ref{sec:overall}). The net smell delta treats all newly introduced smells equally, though some may be less severe than those resolved.

Tool-based measurement of smell resolution may not capture all quality dimensions: a smell that disappears from PyExamine's report may have been resolved through a superficial transformation rather than a genuine architectural improvement. The GEPA prompt optimization pipeline uses a compound proxy metric (static validator + judge model) rather than end-to-end agent performance; validating this proxy against actual outcomes was computationally infeasible and is left to future work. Additionally, GEPA was trained on easy/medium-tier smells from the same codebase; prompts may encode scikit-learn-specific patterns that do not transfer to other projects.

\section{Conclusion}
\label{sec:conclusion}
 
We presented SmellBench, the first benchmark and evaluation framework for LLM-agent-based architectural code smell repair, and evaluated 11 agent configurations from four model families on 65 hard-severity smells in scikit-learn. Three findings stand out. First, the best agent resolves 47.7\% of smells, but four agents converge at a 27.7\% floor, and a substantial portion of resolutions are cascading side effects rather than direct fixes, indicating that deeper architectural reasoning remains a bottleneck. Second, agents identify false positives with up to $\kappa = 0.94$ expert agreement, substantially outperforming their repair capabilities and suggesting that these are separable skills. Third, repair aggressiveness and net codebase quality are inversely related, implying distinct deployment profiles for conservative and aggressive agents. These findings are robust across seven weighting configurations ($W = 0.939$, $p < 0.001$).
 
Our contributions include the SmellBench framework with expert validated ground truth, a composite evaluation methodology that separately assesses repair effectiveness, false positive identification, net codebase impact, as well as the first cross-family agent comparison on architectural smells. Future work should extend the benchmark to additional languages and detection tools, and explore hybrid strategies that combine conservative false positive filtering with scoped repair on confirmed true positives.

\section*{Data Availability}
All experiments and data are available at \url{https://doi.org/10.5281/zenodo.19247588}.

\bibliographystyle{ACM-Reference-Format}
\bibliography{George_ASE_sample-base}

@String{Computing = "Computing" }

@inproceedings{xia2023apr,
author = {Xia, Chunqiu Steven and Wei, Yuxiang and Zhang, Lingming},
    title = {Automated Program Repair in the Era of Large Pre-Trained Language Models},
    year = {2023},
    isbn = {9781665457019},
    publisher = {IEEE Press},
    url = {https://doi.org/10.1109/ICSE48619.2023.00129},
    doi = {10.1109/ICSE48619.2023.00129},
    pages = {1482–1494},
    numpages = {13},
    location = {Melbourne, Victoria, Australia},
    series = {ICSE '23}
}

@inproceedings{bouzenia2025repairagent,
    author = {Bouzenia, Islem and Devanbu, Premkumar and Pradel, Michael},
    title = {RepairAgent: An Autonomous, LLM-Based Agent for Program Repair},
    year = {2025},
    isbn = {9798331505691},
    publisher = {IEEE Press},
    url = {https://doi.org/10.1109/ICSE55347.2025.00157},
    doi = {10.1109/ICSE55347.2025.00157},
    pages = {2188–2200},
    numpages = {13},
    location = {Ottawa, Ontario, Canada},
    series = {ICSE '25}
}

@inproceedings{yang2024sweagent,
    author = {Yang, John and Jimenez, Carlos E. and Wettig, Alexander and Lieret, Kilian and Yao, Shunyu and Narasimhan, Karthik and Press, Ofir},
    title = {SWE-agent: agent-computer interfaces enable automated software engineering},
    year = {2024},
    isbn = {9798331314385},
    publisher = {Curran Associates Inc.},
    address = {Red Hook, NY, USA},
    booktitle = {Proceedings of the 38th International Conference on Neural Information Processing Systems},
    articleno = {1601},
    numpages = {125},
    location = {Vancouver, BC, Canada},
    series = {NIPS '24}
}

@article{xia2025agentless,
    author = {Xia, Chunqiu Steven and Deng, Yinlin and Dunn, Soren and Zhang, Lingming},
    title = {Demystifying LLM-Based Software Engineering Agents},
    year = {2025},
    issue_date = {July 2025},
    publisher = {Association for Computing Machinery},
    address = {New York, NY, USA},
    volume = {2},
    number = {FSE},
    url = {https://doi.org/10.1145/3715754},
    doi = {10.1145/3715754},
    journal = {Proc. ACM Softw. Eng.},
    month = jun,
    articleno = {FSE037},
    numpages = {24},
}

@inproceedings{jimenez2024swebench,
    title={{SWE}-bench: Can Language Models Resolve Real-world Github Issues?},
    author={Carlos E Jimenez and John Yang and Alexander Wettig and Shunyu Yao and Kexin Pei and Ofir Press and Karthik R Narasimhan},
    booktitle={The Twelfth International Conference on Learning Representations},
    year={2024},
    url={https://openreview.net/forum?id=VTF8yNQM66}
}

@inproceedings{wu2024ismell,
    author = {Wu, Di and Mu, Fangwen and Shi, Lin and Guo, Zhaoqiang and Liu, Kui and Zhuang, Weiguang and Zhong, Yuqi and Zhang, Li},
    title = {iSMELL: Assembling LLMs with Expert Toolsets for Code Smell Detection and Refactoring},
    year = {2024},
    isbn = {9798400712487},
    publisher = {Association for Computing Machinery},
    address = {New York, NY, USA},
    url = {https://doi.org/10.1145/3691620.3695508},
    doi = {10.1145/3691620.3695508},
    booktitle = {Proceedings of the 39th IEEE/ACM International Conference on Automated Software Engineering},
    pages = {1345–1357},
    numpages = {13},
    location = {Sacramento, CA, USA},
    series = {ASE '24}
}

@article{cordeiro2024refactoring,
    author    = {Jonathan Cordeiro and Shayan Noei and Ying Zou},
    title     = {An Empirical Study on the Code Refactoring Capability of Large Language Models},
    journal   = {arXiv preprint arXiv:2411.02320},
    url = {https://doi.org/10.48550/arXiv.2411.02320},
    year      = {2024},
}

@article{xu2025mantra,
    author = {Yisen Xu and Feng Lin and Jinqiu Yang and Tse-Hsun (Peter) Chen and Nikolaos Tsantalis},
    title = {MANTRA: Enhancing Automated Method-Level Refactoring with Contextual RAG and Multi-Agent LLM Collaboration},
    journal = {arXiv preprint arXiv:2503.14340v2},
    url = {https://doi.org/10.48550/arXiv.2503.14340},
    year = {2025},
}

@article{oueslati2025refagent,
    author = {Oueslati, Khouloud and Lamothe, Maxime and Khomh, Foutse},
    title = {RefAgent: A Multi-agent LLM-based Framework for Automatic Software Refactoring},
    journal = {arXiv preprint arXiv:2511.03153v2},
    url = {https://doi.org/10.48550/arXiv.2511.03153},
    year = {2026},
}

@article{xue2025smellcc,
author = {Xue, Zhipeng and Zhang, Xiaoting and Gao, Zhipeng and Hu, Xing and Gao, Shan and Xia, Xin and Li, Shanping},
title = {Clean Code, Better Models: Enhancing LLM Performance with Smell-Cleaned Dataset},
year = {2026},
publisher = {Association for Computing Machinery},
address = {New York, NY, USA},
issn = {1049-331X},
url = {https://doi.org/10.1145/3793252},
doi = {10.1145/3793252},
journal = {ACM Trans. Softw. Eng. Methodol.},
month = feb
}

@article{jolak2025maintainability,
    title = {An empirical investigation of the impact of architectural smells on software maintainability},
    journal = {Journal of Systems and Software},
    volume = {225},
    pages = {112382},
    year = {2025},
    issn = {0164-1212},
    doi = {https://doi.org/10.1016/j.jss.2025.112382},
    url = {https://www.sciencedirect.com/science/article/pii/S0164121225000500},
    author = {Rodi Jolak and Simon Karlsson and Felix Dobslaw}
}

@article{shivashankar2025pyexamine,
    author={Shivashankar, Karthik and Martini, Antonio},
    booktitle={2025 IEEE/ACM 22nd International Conference on Mining Software Repositories (MSR)}, 
    title={PyExamine: A Comprehensive, Un-Opinionated Smell Detection Tool for Python}, 
    year={2025},
    volume={},
    number={},
    pages={763-774},
    doi={10.1109/MSR66628.2025.00114}
}

@article{mumtaz2021mapping,
title = {A systematic mapping study on architectural smells detection},
journal = {Journal of Systems and Software},
volume = {173},
pages = {110885},
year = {2021},
issn = {0164-1212},
doi = {https://doi.org/10.1016/j.jss.2020.110885},
url = {https://www.sciencedirect.com/science/article/pii/S0164121220302752},
author = {Haris Mumtaz and Paramvir Singh and Kelly Blincoe},
}

@inbook{tessa2025ecsa,
    author = {Tessa, Claudio and Bochicchio, Matteo and Arcelli Fontana, Francesca},
    year = {2025},
    month = {08},
    pages = {90-98},
    title = {Exploring Architectural Smells Detection Through LLMs},
    isbn = {978-3-032-02137-3},
    doi = {10.1007/978-3-032-02138-0_6}
}

@inproceedings{nursapa2025rose,
    author = {Nursapa, Samal and Samuilova, Anastassiya and Bucaioni, Alessio and Nguyen, Phuong},
    year = {2025},
    month = {10},
    pages = {421-427},
    title = {ROSE: Transformer-Based Refactoring Recommendation for Architectural Smells},
    doi = {10.1109/ESEM64174.2025.00019}
}

@article{guo2023falsepositive,
    author={Guo, Zhaoqiang and Tan, Tingting and Liu, Shiran and Liu, Xutong and Lai, Wei and Yang, Yibiao and Li, Yanhui and Chen, Lin and Dong, Wei and Zhou, Yuming},
    journal={IEEE Transactions on Software Engineering}, 
    title={Mitigating False Positive Static Analysis Warnings: Progress, Challenges, and Opportunities}, 
    year={2023},
    volume={49},
    number={12},
    pages={5154-5188},
    doi={10.1109/TSE.2023.3329667}
}

@article{pedregosa2011scikit,
    author = {Pedregosa, Fabian and Varoquaux, Ga\"{e}l and Gramfort, Alexandre and Michel, Vincent and Thirion, Bertrand and Grisel, Olivier and Blondel, Mathieu and Prettenhofer, Peter and Weiss, Ron and Dubourg, Vincent and Vanderplas, Jake and Passos, Alexandre and Cournapeau, David and Brucher, Matthieu and Perrot, Matthieu and Duchesnay, \'{E}douard},
    title = {Scikit-learn: Machine Learning in Python},
    year = {2011},
    issue_date = {2/1/2011},
    publisher = {JMLR.org},
    volume = {12},
    number = {},
    issn = {1532-4435},
    journal = {J. Mach. Learn. Res.},
    month = nov,
    pages = {2825–2830},
    numpages = {6}
}

@misc{openai2025codex,
    author = {{OpenAI}},
    title = {Codex {CLI}: Open-Source Coding Agent},
    year = {2025},
    url = {https://github.com/openai/codex},
    note = {Accessed: 2026-03-15}
}

@misc{anthropic2025claude,
    author = {{Anthropic}},
    title = {Claude Code: Agentic Coding Tool},
    year = {2025},
    url = {https://github.com/anthropics/claude-code},
    note = {Accessed: 2026-03-15}
}

@misc{google2025gemini,
    author = {{Google}},
    title = {Gemini {CLI}: {AI} Agent for the Command Line},
    year = {2025},
    url = {https://github.com/google-gemini/gemini-cli},
    note = {Accessed: 2026-03-15}
}

@misc{mistral2025vibe,
    author = {{Mistral AI}},
    title = {Mistral Vibe: Agentic Coding Assistant},
    year = {2025},
    url = {https://github.com/mistralai/mistral-vibe},
    note = {Accessed: 2026-03-15}
}

@article{agrawal2025gepa,
    title = {GEPA: Reflective Prompt Evolution Can Outperform Reinforcement Learning},
    author = {Lakshya A. Agrawal and Shangyin Tan and Dilara Soylu and Noah Ziems and Rishi Khare and Krista Opsahl-Ong and Arnav Singhvi and Herumb Shandilya and Michael J. Ryan and Meng Jiang and Christopher Potts and Koushik Sen and Alexandros G. Dimakis and Ion Stoica and Dan Klein and Matei Zaharia and Omar Khattab},
    journal = {arXiv preprint arXiv:2507.19457v2},
    year = {2026},
    url = {https://doi.org/10.48550/arXiv.2507.19457}
}

@inproceedings{arcuri2011practical,
author = {Arcuri, Andrea and Briand, Lionel},
title = {A practical guide for using statistical tests to assess randomized algorithms in software engineering},
year = {2011},
isbn = {9781450304450},
publisher = {Association for Computing Machinery},
address = {New York, NY, USA},
url = {https://doi.org/10.1145/1985793.1985795},
doi = {10.1145/1985793.1985795},
booktitle = {Proceedings of the 33rd International Conference on Software Engineering},
pages = {1–10},
numpages = {10},
location = {Waikiki, Honolulu, HI, USA},
series = {ICSE '11}
}

@inproceedings{batole2025localizeagent,
author = {Batole, Fraol and OBrien, David and Nguyen, Tien N. and Dyer, Robert and Rajan, Hridesh},
title = {An LLM-Based Agent-Oriented Approach for Automated Code Design Issue Localization},
year = {2025},
isbn = {9798331505691},
publisher = {IEEE Press},
url = {https://doi.org/10.1109/ICSE55347.2025.00100},
doi = {10.1109/ICSE55347.2025.00100},
booktitle = {Proceedings of the IEEE/ACM 47th International Conference on Software Engineering},
pages = {1320–1332},
numpages = {13},
location = {Ottawa, Ontario, Canada},
series = {ICSE '25}
}

@article{velasco2026causal,
    author = {Alejandro Velasco and Daniel Rodriguez-Cardenas and Dipin Khati and David N. Palacio and Luftar Rahman Alif and Denys Poshyvanyk},
    title = {A Causal Perspective on Measuring, Explaining and Mitigating Smells in LLM-Generated Code},
    journal   = {arXiv preprint arXiv:2511.15817v5},
    url = {https://doi.org/10.48550/arXiv.2511.15817},
    year = {2026}
}

@misc{anthropic2024mcp,
  author       = {{Anthropic}},
  title        = {Introducing the Model Context Protocol},
  year         = {2024},
  month        = nov,
  howpublished = {Anthropic Blog},
  url          = {https://www.anthropic.com/news/model-context-protocol},
  note         = {Accessed: 2026-03-21}
}

@article{suryanarayana2014refactoring,
author = {Tracz, Will},
title = {Refactoring for Software Design Smells: Managing Technical Debt by Girish Suryanarayana, Ganesh Samarthyam, and Tushar Sharma},
year = {2015},
issue_date = {November 2015},
publisher = {Association for Computing Machinery},
address = {New York, NY, USA},
volume = {40},
number = {6},
issn = {0163-5948},
url = {https://doi.org/10.1145/2830719.2830739},
doi = {10.1145/2830719.2830739},
journal = {SIGSOFT Softw. Eng. Notes},
month = nov,
pages = {36},
numpages = {1}
}

@article{gautam2025refactorbench,
  author={Dhruv Gautam and Spandan Garg and Jinu Jang and Neel Sundaresan and Roshanak Zilouchian Moghaddam},
  title={RefactorBench: Evaluating Stateful Reasoning in Language Agents Through Code},
  journal   = {arXiv preprint arXiv:2503.07832v1},
  url = {https://doi.org/10.48550/arXiv.2503.07832},
  year={2025}
}

\end{document}